\documentclass[aps,prd,reprint,superscriptaddress,nofootinbib]{revtex4-1}
\usepackage{cuted} 
\usepackage[normalem]{ulem}
\usepackage[normalem]{ulem}
\usepackage{xcolor}
\usepackage{mdframed}
\usepackage[T1]{fontenc} 

\usepackage{graphicx,epsfig}
\usepackage{bbm}
\usepackage{graphics,dcolumn,bm, float}
\usepackage{amssymb,amsmath,rotate,color,amsfonts}
\usepackage[title,titletoc,toc]{appendix}
\usepackage{mathtools}
\usepackage{booktabs}
\usepackage{tcolorbox}
\usepackage{tikz-cd}
\usetikzlibrary{shadings}
\usepackage{xcolor}
\usepackage{tikz-3dplot}
\usepackage{setspace}
\usepackage{wrapfig}
\usepackage{lipsum}
\usepackage{mwe}
\usepackage[mathlines]{lineno}
\usepackage[mathscr]{euscript}
\usepackage{bbold}
\usepackage{pgfplots}
\usepackage{float}
\usepackage{tkz-euclide}
\usepackage{braket}
\usepackage{physics}
\usepackage[caption=false]{subfig}
\usepackage[export]{adjustbox}
\usepackage{tikz}
\usepackage{slashed}
\usepackage{bm}
\usepackage{multirow}
\usetikzlibrary{through,calc}
\usetikzlibrary{positioning}

\usepackage{hyperref}

\makeatletter
\def\pre@bibdata{revtex-notes}
\makeatother

\newcommand{\be}{\begin{equation}}
	\newcommand{\ee}{\end{equation}}
\newcommand{\bea}{\begin{eqnarray}}
	\newcommand{\eea}{\end{eqnarray}}
\newcommand{\nn}{\nonumber}
\newtheorem{definition}{Definition}[section]

\tdplotsetmaincoords{70}{120} 

\begin{document}

\nocite{*}

\title{$BMS_3$-like algebras via the $Z_N$-graded $u(1)^2$ Kac-Moody algebra }

\preprint{}

\author{Armin Ghazi}
\email{armin.gh.kh@gmail.com}
\author{Ahmad Moradpouri}
\email{Ahmadreza.Moradpour@gmail.com}
\affiliation{Research Center for High Energy Physics, Department of Physics, Sharif University of Technology, P.O.Box 11155-9161, Tehran, Iran.}
\begin{abstract}
	The Sugawara construction provides a natural way to construct the Virasoro algebra from a current algebra. It was shown in Ref.~\cite{Ghazi:2025oin} that for the $u(1)^2$ Kac--Moody current algebra, there exist additional constructions that exhibit a $\mathbb{Z}_N$-graded structure. Indeed, the space of such constructions defines a non-compact algebraic variety whose dimension depends on $N$. In this paper, we consider the compactification of these algebraic varieties by adding points at infinity to the non-compact part, and show that these points correspond precisely to generalizations of $BMS_3$-like algebras. More explicitly, for a $\mathbb{Z}_2$ grading, the corresponding algebra coincides with the $BMS_3$ algebra, which takes the form $\mathrm{Vir} \rtimes F$, where $F$ is an infinite abelian ideal of the full algebra. For $N > 2$, we show that there exist generalizations of the standard $BMS_3$ algebra of the form $\mathrm{Vir} \rtimes F$, where $F$ is a nonabelian ideal that forms a nilpotent algebra of depth $r < N$. We further demonstrate that the depth of the algebra is related to the order of the singularity of the algebraic variety at that point.We also show that the polynomials defining the algebraic varieties exhibit a factorization property into linear factors, which, if true, classifies all $BMS_3$-like algebras. Finally, we study the central extensions of these algebras, which are consistent with the general structure of algebras corresponding to primary fields of conformal weight $h = 2$. 
\end{abstract}

	\maketitle
	\flushbottom
	
	
	
	
	\section{Introduction}
Two-dimensional conformal field theory (CFT) has played a central role in theoretical physics over the past several decades. In their seminal work, Belavin, Polyakov, and Zamolodchikov~\cite{BELAVIN1984333} established the foundations of two-dimensional CFT, which has since found numerous applications ranging from pure mathematics to condensed matter physics. For a detailed exposition from the physical perspective, see~\cite{DiFrancesco:1997nk,ginsparg1988appliedconformalfieldtheory,Blumenhagen:2009zz}, and for a mathematical treatment, see~\cite{Schottenloher:2008zz}. It plays a significant role in the study of the two-dimensional exactly solvable models, for example~\cite{Friedan1986NotesOS,Polyakov:1970xd,Friedan:1984rv}. In statistical mechanics, two-dimensional CFTs have proven to be powerful in the study of critical phenomena, stochastic evolutions, and universal features of entanglement entropy~\cite{CARDY1986186,cardy2008conformalfieldtheorystatistical,Itzykson:1989sy,Bauer_2006,calabrese2005entanglemententropyquantumfield,Cardy_2005}. It also finds many applications in condensed matter physics, ranging from the study of critical models such as the Ising and Potts models to the description of the fractional quantum Hall effect~\cite{tong2016lecturesquantumhalleffect,fradkin2021quantum,MOORE1991362,Fateev:1985mm}. Two-dimensional CFTs also lie at the heart of research in string theory~\cite{Polchinski:1998rq,Becker_Becker_Schwarz_2006,Blumenhagen:2013fgp}. From a pure mathematical point of view, it has a significant impact on vertex operator algebras~\cite{RichardBorcherds,borcherds1999quantumvertexalgebras,Frenkel1988xz}, representation of affine Lie algberas~\cite{kac1990infinite,MOODY1968211,Lepowsky:1978jk,frenkel1992vertex,Frenkel:1980rn,Goddard:1986ee,kac1987bombay,FeiginFuchs1982}. Two-dimensional CFTs also make a broad contribution to research in mathematical physics~\cite{VERLINDE1988360,Moore:1988qv,Moore:1989yh,frenkel2005lectureslanglandsprogramconformal,Goddard:1986bp,Dixon:1985jw,DixonFriedanMartinecShenker1987,DijkgraafVafaVerlindeVerlinde1989}. 

The Virasoro algebra~\cite{DiFrancesco:1997nk,ginsparg1988appliedconformalfieldtheory,Blumenhagen:2009zz},
\bea
\label{Virasoroalgebra}
[L_n,L_m]=(n-m)L_{n+m}+\frac{c}{12}n(n^2-1)\delta_{n+m,0},\eea
which is the central extension of the Witt algebra, serves as the symmetry algebra of two-dimensional conformal symmetry. A substantial part of the research in this area focuses on the classification of unitary representations of the Virasoro algebra~\cite{BELAVIN1984333,Felder:1989wr,Goddard:1986bp,Goddard:1986ee,Kac:1979fz,kac1987bombay, Friedan:1984rv,Friedan:1986ua}. The Virasoro algebra can also be extended into Kac-Moody-Virasoro algebras, where the Kac-Moody algebra~\cite{kac1990infinite,MOODY1968211,kac1987bombay}, $\hat g_k$ for Lie algebra $g$ and level $k$ is given by
\bea
\label{kacmoody}
[j^A_m, j^B_n] = i f^{AB}_{\;\;C} j^C_{m+n} + k\, m \, \delta^{AB} \delta_{m+n,0}.
\eea
In this broader context, the Virasoro generators are not unrelated to the Kac-Moody generators, and the Sugawara~\cite{PhysRev.170.1659} construction suggests a natural way of how $L_n$ can be constructed in terms of the Kac-Moody generators $j_n^A$ as follows
\bea
T(z)=\gamma\sum_{A=1}^{dim~g}N(J^AJ^A)(z)\eea
where $J^{A}=\sum_n j^A_nz^{-n-1}$ and $T(z)=\sum L_n z^{-n-2}$ for an appropriate $\gamma$. However, the Sugawara construction does not represent the most general expression of the Virasoro generators in terms of current modes, and various extensions of this construction have been explored in the literature. The general Virasoro affine construction on an affine Lie algebra $\mathfrak{g}$ was studied by M.B. Halpern and E. Kritsis~\cite{Halpern:1989ss}, and further explored in subsequent works~\cite{HALPERN1991333,HALPERN1990411,Halpern:1989zy,de_Boer_1997}.  In~\cite{Ghazi:2025oin}, the $\mathbb{Z}_N$-equivariant Virasoro algebra associated with the group $U(1)^2$ was introduced, providing a framework for constructing various realizations of the Virasoro generators in terms of the current modes $j^A_n$. As the present paper builds upon the results of Ref.~\cite{Ghazi:2025oin} and on the Virasoro algebra via the $Z_{N}$-equivariant Virasoro algebra, it is necessary to summarize the key findings of Ref.~\cite{Ghazi:2025oin}.

The general structure of the $\mathbb{Z}_N$-equivariant construction is as follows. The Virasoro generators $L_n$ and the current modes $j^A_n$ are decomposed into sets of generators ${L^k_n}$ and ${j^{kA}_n}$, where $k = 0, 1, \ldots, N - 1$. Here, $L^k_n$ (respectively $j^{kA}_n$) denotes the component for which the mode index satisfies $n \in N\mathbb{Z} + k$.
The main question addressed in~\cite{Ghazi:2025oin} was the following: given a fixed Kac–Moody algebra~\eqref{kacmoody}, what is the most general construction of the Virasoro generators in terms of the current modes for $g = u(1)^D$?
The answer for $D=2$ is rather surprising. In $D=2$, in addition to the standard delta tensor $\delta^{AB}$,  there exists another rotationally invariant tensor, namely the epsilon tensor $\epsilon^{AB}$, which allows for the definition of new operators beyond the standard Sugawara construction~\eqref{kacmoody}
\bea
\label{newoperator}
S_n=\epsilon^{AB}\sum_q f(q)j^{A}_{n-q}j^B_q.
\eea

\begin{figure}[h]
	\centering
	\includegraphics[width=0.5\textwidth]{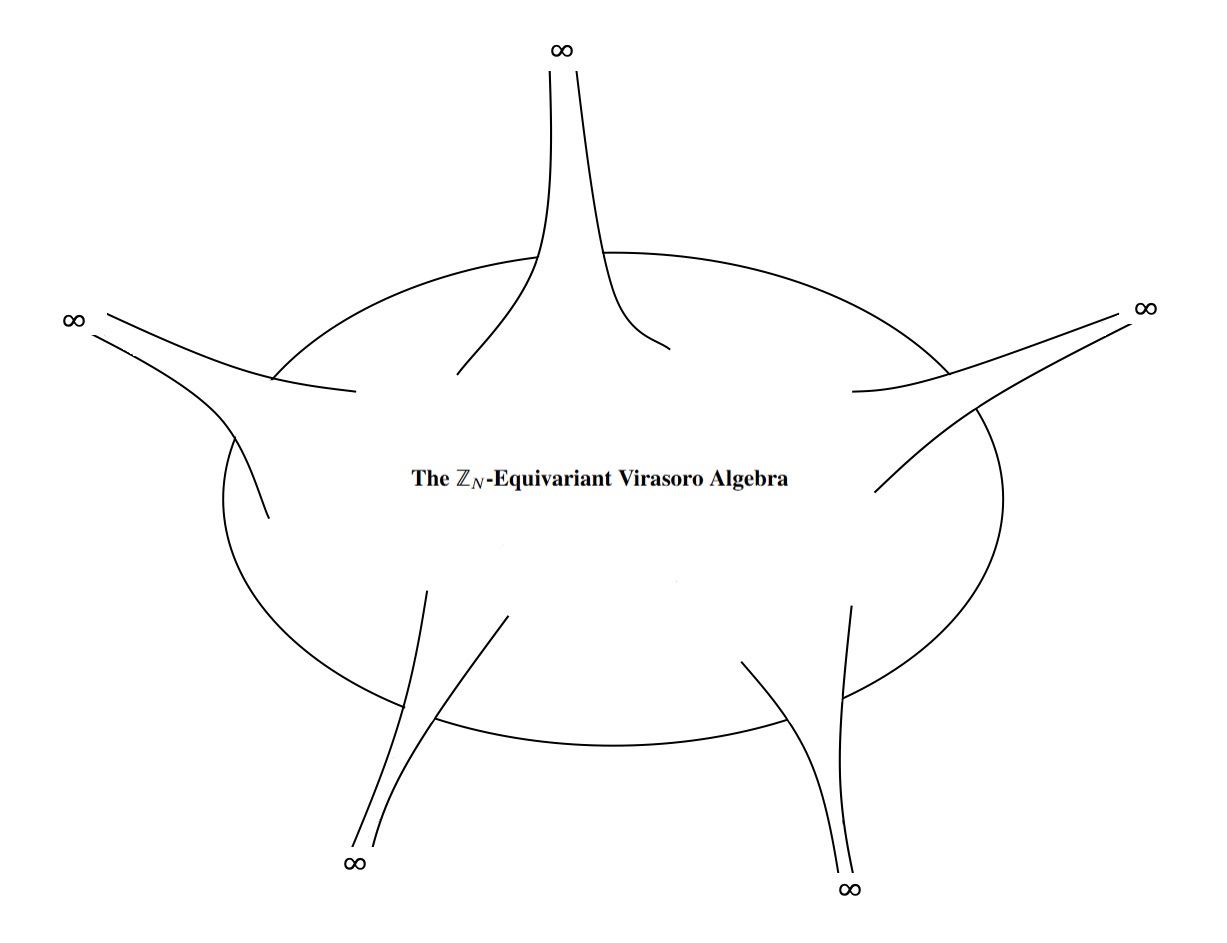}
	\caption{Non-compact algebraic variety corresponding to the $\mathbb{Z}_N$ equivariant construction of the Virasoro algebra in terms of current modes $j^A_n$.}
	\label{ZNEquivariant}
\end{figure}

Therefore, one can construct more general linear combinations of these operators as follows:
\bea
\label{newconstruction}
\sim\alpha_i\delta^{AB}\sum_q g_i(q)j^{A}_{n-q}j^B_q+\beta_i\epsilon^{AB}\sum_q f_i(q)j^{A}_{n-q}j^B_q+...,
\eea
where $\alpha_i$ and $\beta_i$ are complex numbers, and $g_i(q)$ and $f_i(q)$ are appropriate functions. The requirement that the Virasoro algebra closes fully determines the functions $g_i(q)$ and $f_i(q)$ imposes an algebraic constraint on the coefficients $\alpha_1$ and $\beta_i$ of the form
\bea
\label{polynomial}
\mathcal{F}(\alpha_i,\beta_i)=0
\eea
where $\mathcal{F}$ is a polynomial function which simply defines a variety. It is shown in~\cite{Ghazi:2025oin} that the space of such constructions for the $\mathbb{Z}_N$-equivariant Virasoro algebra is a non-compact variety where its dimensions depends on $N$ which is schematically shown in Fig~\ref{ZNEquivariant}.  Therefore, there exist points at infinity such that, by adjoining them to the algebraic variety, one obtains a compact projective variety embedded in $\mathbb{C}P^N$.

The purpose of this paper is to investigate the structure of these points at infinity. By definition, they do not correspond to the Virasoro algebra; otherwise, they would lie on the non-compact part of the algebraic variety, which would contradict the very fact that they represent points at infinity.

The main results of this paper consist of two parts. First, the variety defined by the constraints takes the form
\bea
(x_1x_2...x_N=1)/\sim
\eea
where $x_i$, for $i=1,...,N$, are linear functions of $\alpha_i,\beta_i$. The equivalence relation $\sim$ arises from the automorphism group of the $u(1)^2$ Kac-Moody algebra that preserve the $Z_N$-grading
\bea
j^{kA}_n\to \lambda_kj^{kA}_n,~~~j^{(N-k)A}_n\to \lambda^{-1}_kj^{(N-k)A}_n
\eea
for $\lambda_k\neq0$. This automorphism group imposes an equivalence relation on the $x_i$, which must be taken into account to identify the true space of inequivalent Virasoro constructions arising from the current algebra. However, in this note, we will disregard this equivalence relation and focus on the simpler equation $x_1x_2...x_N=1$. The effect of the equivalence relation is merely to rearrange the coefficients of the general construction~\eqref{newconstruction}. 
 
The corresponding projective variety, which can be constructed by setting $x_i=\frac{X_i}{Z}$, takes the following form up to equivalence:

\bea
\label{facconjecture1}
Z^{N} = X_{1} X_{2} \cdots X_{N},
\eea
 where the points at infinity correspond to $Z=0$. Consequently, the points at infinity lie on the intersection of the hyperplanes $X_i=0$. For a given point $\tilde x$ at infinity which is typically a singular point of the projective variety, depending on how many lines it lies on them, the resulting algebra depends on how many of these hyperplanes the point lies on. Determining this dependence constitutes the second main finding of this paper. Specifically, the algebras associated with the points at infinity are, in general, of the form
\bea
g=Vir\rtimes F,
\eea
where $F$ is an infinite-dimensional nilpotent Lie algebra of depth $\mathcal{K} < N \in \mathbb{N}$ for a given $\mathbb{Z}_N$. The depth of the algebra $F$ is related to the order of the singularity of the corresponding point at infinity. We explicitly verify these results for $\mathbb{Z}_2$, $\mathbb{Z}_3$, and $\mathbb{Z}_4$, while leaving the general proof for future work. In particular, for $\mathbb{Z}_2$ it is shown that the corresponding algebra is closely related to the $BMS_3$ algebra, where $F$ is simply an abelian ideal. For $N = 3, 4$, the structure includes extension of the Virasoro algebra by non-abelian ideals which can be considered as analogues of the $BMS_3$ algebra. 

The organization of this paper is as follows. In Section~\ref{ZNequivariant}, we summarize the essential aspects of the $\mathbb{Z}_N$-equivariant Virasoro algebra and discuss how alternative Sugawara-type constructions naturally emerge within this framework. In Section~\ref{Factorizationconjecture}, we present the two main conjectures of the paper, the factorization conjecture and the emergence of nilpotent algebras and also indicate that the factorization conjecture completely classifies the nature of points at infinity.  Section~\ref{closure} constitutes the main part of this paper, where we demonstrate that both conjectures hold for $\mathbb{Z}_2$, $\mathbb{Z}_3$, and $\mathbb{Z}_4$. In particular, we show that the points at infinity for $\mathbb{Z}_2$ consist of two isolated points, and the corresponding algebras represent truncations of the $BMS_3$ or conformal Galilean algebra. We therefore refer to them as half Conformal Galilean Algebras (hCGA). For $\mathbb{Z}_3$, the structure is richer and includes higher-order or singular points. More generally, we show that the corresponding algebra is isomorphic to $Vir \rtimes F$, where $F$ is a nilpotent algebra of depth $\mathcal{K} = N-r$, with $r$ denoting the order of the singularity of the point. In Section~\ref{newalgebras}, we relax the $\mathbb{Z}_N$-equivariance condition and study these algebras from a purely algebraic perspective, focusing on their possible central extensions. To the best of the authors' knowledge, these algebras have not been discussed in the existing literature and may therefore be regarded as new structures worthy of independent consideration.

\section{The $\mathbb{Z}_N$-Equivariant Virasoro Algebra}
\label{ZNequivariant}

The Virasoro or Witt algebra are good examples of $\mathbb{Z}_N$-graded algebras. Consider the Witt algebra
\bea
[\ell_n, \ell_m] = (m - n)\, \ell_{m+n},
\eea
which clearly exhibits such a grading. The Witt generators $\ell_n$ decompose as
\bea
\ell_{n}^k,
\eea
for $k = 0, 1, \dots, N-1$ and $[n] = k \mod N$, and the Witt algebra respects this grading as follows
\bea
[\ell_n^{k_1}, \ell_m^{k_2}] = (m - n)\, \ell_{m+n}^{k_1+k_2},
\eea

From the Lagrangian point of view, consider a free boson theory on a flat cylinder parametrized by $(\tau, \theta)$. The left-moving part of the energy-momentum tensor is given by
\begin{equation}
	T_{++} \sim \partial_{+}X \partial_{+}X = \sum_n \ell_n e^{in\theta^{+}} = \sum_{k=0}^{N-1} \; \sum_{[n]=k} \ell_{n}^k e^{in\theta^{+}},
\end{equation}
where $\theta^{+} = \tau + \theta$. Let us denote $\partial_{+}X^k = \sum_{[n]=k} j_n e^{in\theta^{+}}$. Then it is evident that the coefficients $\ell_{n}^k$ arise from products of the form
\bea
\partial_{+}X^{k_1} \partial_{+}X^{k_2},
\eea
with $k_1 + k_2 = k$.

This grading can be systematically achieved from the $\mathbb{Z}_N$-equivariance concept. Explicitly, let $S^1$ denote the unit circle, parameterized by an angle $\theta \in [0,2\pi)$.  
The group of smooth orientation-preserving diffeomorphisms is

\begin{equation}
	\mathrm{Diff}^+(S^1) = \bigl\{ f: S^1 \to S^1 \;\big|\; f \text{ is smooth, bijective, and } f'(\theta) > 0 \bigr\},
\end{equation}
and the Lie algebra of $Diff^+(S^1)$ consists of smooth vector fields on the circle:
\bea
Vect(S^1) = \{ v(\theta)\partial_\theta \, | \, v(\theta) \in C^\infty(S^1) \}.\eea
The Lie bracket is given by the usual commutator of vector fields:
\bea
\label{vectorfields}
[v_1(\theta)\partial_\theta, v_2(\theta)\partial_\theta] = 
\big( v_1 v_2' - v_2 v_1' \big)\partial_\theta,
\eea
This algebra captures the infinitesimal generators of circle reparametrizations.

Now suppose we have a homomorphism $\psi:\mathbb{Z}_n\to Diff(S^1)$. A function $f_k(\theta)$ is said to be $\mathbb{Z}_N$-equivariant with respect to the one-dimensional irreducible representation $\rho_k(g)=e^{\frac{2\pi i k}{N}}$ of $\mathbb{Z}_N$, (where $g$ is the generator) if it satisfies the equivariance condition
\bea
f_k(\psi_g(\theta)) = \rho_k(g)\,f_k(\theta),
\qquad \forall \theta \in S^1,
\eea
which can be shown by the following commutative diagram
\bea
\begin{tikzcd}
	S^1 \arrow[r, "\psi_g"] \arrow[d, "f_k"'] & 
	S^1 \arrow[d, "f_k"] \\[2pt]
	\mathbb{C} \arrow[r, "\rho_k(g)"'] & 
	\mathbb{C}
\end{tikzcd}
\eea
The above equivariance property decompose the space of functions $C^{\infty}(S^1,\mathbb{C})$ into these isotypic subspaces which is obtained by means of natural projection operators. Specifically, the projection onto $C^{\infty}(M,\mathbb{C})_{\rho}$ can be expressed by averaging over the group $G$ using the character $\chi_{\rho}$ of $\rho$,
\bea
(P_{\rho} f)(\theta) \;:=\; \frac{d_\rho}{|G|} \sum_{g \in G} \chi_\rho(g^{-1}) \, f(\psi_{g}(\theta)),
\eea
where $d_\rho = \dim \rho$ and $\chi_\rho = \mathrm{Tr}(\rho)$ denotes the character of $\rho$. Since $G$ is finite and Abelian, every irreducible representation $\rho$ is one-dimensional, so $d_\rho = 1$ and $\chi_\rho(g) = \rho(g)$ for all $g \in G$.

The standard orthogonality relations of characters imply the following properties:
\begin{equation}
	P_{\rho}^{2} = P_{\rho}, \qquad 
	P_{\rho} P_{\sigma} = 0 \;\; \text{if } \rho \not\simeq \sigma, \qquad
	\sum_{\rho \in \widehat{G}} P_{\rho} = \mathrm{Id}_{C^{\infty}(M)}.
\end{equation}
Here $\widehat G$ denotes the set of all inequivalent irreducible representations of $G$. Therefore, the projection operators $\{P_\rho\}_{\rho \in \widehat{G}}$ forms a complete set of mutually orthogonal projections, and one obtains the direct sum decomposition:
\bea
\label{decomposition1}
C^{\infty}(S^1,\mathbb{C}) \;\cong\; \bigoplus_{\rho \in \widehat{G}} C^{\infty}(S^1,\mathbb{C})_{\rho}.
\eea
Such a decomposition also induces a $\mathbb{Z}_N$-graded algebra structure on the space of functions. Concretely, if 
$f_{k_1}\in C^{\infty}(S^1,\mathbb{C})_{k_1}$ and $f_{k_2}\in C^{\infty}(S^1,\mathbb{C})_{k_2}$, then their pointwise product satisfies
\bea
f_{k_1}(p)\cdot f_{k_2}(p)\;\in\;C^{\infty}(S^1,\mathbb{C})_{k_1+k_2\pmod N}.
\eea
Thus, the grading is additive modulo $N$. 

Interestingly, this $\mathbb{Z}_N$-grading also applies to the commutators of the vector fields~\eqref{vectorfields}. Concretely, if 
$v_1\in C^{\infty}(S^1,\mathbb{C})_{k_1}$ and $v_2\in C^{\infty}(S^1,\mathbb{C})_{k_2}$, 
then their Lie bracket
\begin{equation}
	[v_1, v_2] = v_1 v_2' - v_2 v_1' \;\in\; C^{\infty}(S^1,\mathbb{C})_{k_1+k_2\pmod{N}}.
\end{equation}
Hence, the space of vector fields exhibits a natural $\mathbb{Z}_N$-graded Lie algebra structure, with the grading additive modulo $N$. Equivalently, the Witt algebra in its $\mathbb{Z}_N$-graded form reads
\bea
[\ell_n^{k_1}, \ell_m^{k_2}] = (m - n)\, \ell_{m+n}^{k_1+k_2},
\eea
where $\ell_n^{k}$ show that the $n\in N\mathbb{Z}+k$ for $k=0,1,...,N-1$. Trivially, such $\mathbb{Z}_N$-gradding is also true for the Virasoro algebra.

The $\mathbb{Z}_N$-graded structure described above may appear rather superficial if one considers the Virasoro algebra in isolation. However, for the $u(1)^2$ Kac--Moody algebra, the existence of an additional invariant tensor, namely the epsilon tensor, as explained in Ref.~\cite{Ghazi:2025oin}, when combined with the concept of $\mathbb{Z}_N$-equivariance, allows for the construction of new realizations beyond the standard one, as we discuss in the next section.
	
\subsection{The Alternative Sugawara Constructions for $\mathbb{Z}_N$–-Equivariant Virasoro Algebras}

As it is explained in the introduction, the aim of the paper~\cite{Ghazi:2025oin} was to propose alternative constructions of the energy--momentum tensor for abelian setting. Specifically, by considering the abelian algebra $\mathfrak{g} = u(1)^{2}$ and setting the level to $k=1$, we decompose the currents into their $\mathbb{Z}_{N}$--equivariant components as
\begin{equation}
	j^{kA}_{n} \equiv j^{A}_{n}, \qquad \text{when } n \in N\mathbb{Z} + k, \quad k = 0,1,\ldots,N-1.
\end{equation}
On this basis, the Kac–Moody algebra for the abelian case takes the form
\bea
\label{equivariantcurrent}
\big[ j^{k_{1}A}_{n}, j^{k_{2}B}_{m} \big]
= n\, \delta^{AB}\, \delta^{k_{1}+k_{2}=0}\, \delta_{m+n,0},
\eea
where the condition $k_{1}+k_{2}=0$ is understood modulo $N$. The currents $j^{kA}_n$ have a natural $SO(2)$ rotation on the index $A$. In $D=2$, there exists another invariant tensor  
\bea
\epsilon^{AB},\nn
\eea
which allows us to construct a more general form for the Virasoro generators. The most general\footnote{For the most general term, we mean that it can be written entirely in terms of the modes $j^{A}_{n}$, while keeping $L^{0}_{n}$ fixed as the standard construction proposed in~\eqref{standardconstruction}.
} form of the $\mathbb{Z}_N$--equivariant set of Virasoro generators $L_n^{k}$, expressed in terms of the current modes $j^{A}_{n}$,\footnote{When we write $j^{A}_{n}$, the index $n$ runs over all integers. However, in the notation $j^{kA}_{n}$, the index $n$ is restricted to $n \in N\mathbb{Z} + k$.} takes the following form~\cite{Ghazi:2025oin}:
\begin{equation}
	\label{Sdefinition}
	L^{k}_{n}
	= \frac{1}{2}\sum_{q\in\mathbb{Z}}\sum_{w=0}^{N-1}
	\Bigl(
	d^{k}_{w}\, e^{2\pi i w q/N}\, j^{A}_{n-q} j^{A}_{q}
	+ c^{k}_{w}\, e^{2\pi i w q/N}\, \epsilon^{AB} j^{A}_{n-q} j^{B}_{q}
	\Bigr),
\end{equation}
where $k=1,\ldots, N-1$ and $n\in N\mathbb{Z}+k$. The coefficients $d^{k}_{w}$ and $c^{k}_{w}$ are determined by demanding the consistency of the algebra and the closure of the Virasoro generators under commutation. These coefficients must satisfy the following constraints:
\bea
\label{cons1}
d^{k}_{N-w} = d^{k}_{w}\, e^{\frac{2\pi i w k}{N}}, 
\qquad
c^{k}_{N-w} = -\, c^{k}_{w}\, e^{\frac{2\pi i w k}{N}}.
\eea

The first term in~\eqref{Sdefinition} corresponds to the symmetric contraction of the current modes by $\delta^{AB}$, while the second term, involving $\epsilon^{AB}$, captures an antisymmetric mixing between different abelian components of the currents. The interplay between these two types of contractions allows one to construct nontrivial $\mathbb{Z}_N$--equivariant extensions of the Sugawara form. For the sector ($k=0$), we assume the standard Sugawara construction:
\bea
\label{standardconstruction}
L^{0}_{n} = \frac{1}{2}\sum_{q} j^{A}_{n-q} j^{A}_{q},
\qquad n\in N\mathbb{Z}.
\eea
This provides the reference point from which the sectors ($k\neq0$) generalize the definition of the Virasoro generators. The consistency of these definitions ensures that the full set $\{L^{k}_{n}\}$ satisfies a closed algebra compatible with the $\mathbb{Z}_N$ symmetry.

This decomposition provides a natural setting for constructing $\mathbb{Z}_{N}$--equivariant versions of the Sugawara energy--momentum tensor, potentially leading to new classes of conformal field theories beyond the standard abelian case. The closedness of the Witt algebra, $[L^{k_1}_n,L^{k_1}_m]=(n-m)L^{k_1+k_2}_{n+m}$, leads to the following relations.\footnote{In the computations, we ignore normal ordering, which becomes important when $k_1 + k_2 \equiv 0 \mod N$. The central term of the Virasoro algebra can be determined consistently via the Jacobi identity.}
\begin{align}
	\label{eq1}
	\sum_{\substack{w_1,w_2 \\ w_1+w_2=s \mod N}} 
	&\bigl( d^{k_1}_{w_1} d^{k_2}_{w_2} - c^{k_1}_{w_1} c^{k_2}_{w_2} \bigr)
	\left( \frac{\zeta_{k_1}^{-w_2} + \zeta_{k_2}^{-w_1}}{2} \right) \nonumber \\
	&\qquad = d^{k_1+k_2}_s \\[1ex]
	\label{eq2}
	\sum_{\substack{w_1,w_2 \\ w_1+w_2=s \mod N}} 
	&\bigl( d^{k_1}_{w_1} c^{k_2}_{w_2} + c^{k_1}_{w_1} d^{k_2}_{w_2} \bigr)
	\left( \frac{\zeta_{k_1}^{-w_2} + \zeta_{k_2}^{-w_1}}{2} \right) \nonumber \\
	&\qquad = c^{k_1+k_2}_s
\end{align}
where $\zeta_k=e^{\frac{2\pi i k}{N}}$. Furthermore, the following constraints have to be satisfied by $c^k_{\omega}$ and $d^k_{\omega}$:
\begin{equation}
	\label{eq3}
	\sum_{\substack{w_1,w_2 \\ w_1+w_2=s \mod N}} 
	\bigl( d^{k_1}_{w_1} d^{k_2}_{w_2} - c^{k_1}_{w_1} c^{k_2}_{w_2} \bigr)
	\Bigl( \frac{\zeta_{k_1}^{-w_2} - \zeta_{k_2}^{-w_1}}{2} \Bigr) = 0
\end{equation}
\begin{equation}
	\label{eq4}
	\sum_{\substack{w_1,w_2 \\ w_1+w_2=s \mod N}} 
	\bigl( d^{k_1}_{w_1} c^{k_2}_{w_2} + c^{k_1}_{w_1} d^{k_2}_{w_2} \bigr)
	\Bigl( \frac{\zeta_{k_1}^{-w_2} - \zeta_{k_2}^{-w_1}}{2} \Bigr) = 0
\end{equation}
for all $s = 0, 1, \ldots, N-1$.

In the next section, we identify the space of all solutions that are consistent with the $\mathbb{Z}_N$ equivariance for $N=2,3$ and $N=4$.	

	\section{Factorization conjecture}
\label{Factorizationconjecture}
The space of $\mathbb{Z}_N$--equivariant Virasoro algebras is not unique, and a detailed investigation of these algebras for $N=2$, $N=3$, and $N=4$ has been carried out in~\cite{Ghazi:2025oin}. As it is noted

One of the main goals of this paper consists of providing supporting evidence for a conjecture, referred to as the \emph{factorization conjecture}, which states that

\medskip
\noindent\textbf{Factorization Conjecture.} 
\label{factorization}
The projective variety corresponding to the different $\mathbb{Z}_{N}$-equivariant Virasoro algebra constructions throught the current modes $j^A_n$ can be factorized into linear component $X_i$ and embedded into $\mathbb{CP}^{N}$ as
\bea
\label{facconjecture}
Z^{N} = X_{1} X_{2} \cdots X_{N},
\eea
for some suitable choice of homogeneous coordinates on $\mathbb{CP}^{N}$ which will be introduced in the next examples.

We do not prove the above conjecture for arbitrary $N$; however, we demonstrate its validity for $\mathbb{Z}_2$, $\mathbb{Z}_3$, and $\mathbb{Z}_4$, which provides supporting evidence for its correctness in the general case. As explained in the introduction, the space of all inequivalent $\mathbb{Z}_{N}$-equivariant Virasoro algebras corresponds to the set of points in the projective patch defined by $Z \neq 0$, which is represented by the coordinate chart
\bea
[X_{1}:\cdots:X_{N}:Z] \;\longmapsto\; \Big(\frac{X_{1}}{Z},\, \ldots,\, \frac{X_{N}}{Z},\, 1\Big),
\eea
this chart defines the following affine hypersurface in $\mathbb{C}^{N}$:
\bea
\label{affine}
x_{1}x_{2}\cdots x_{N} = 1,
\eea
where $x_i=\frac{X_i}{Z}$ and in general, the coordinates $x_{i}$ ($i=1,\ldots,N$) are linear functions of the coefficients $d^{k}_{\omega}$ and $c^{k}_{\omega}$ appearing in the construction~\eqref{Sdefinition}.
\subsection{Consequences of the Factorization Conjecture: Emergence of Nilpotent Algebras}

Our focus is to investigate the algebras associated with the points at infinity of the affine curve~\eqref{affine}, which are defined by the condition $Z = 0$ in equation~\eqref{facconjecture}. We refer to the space of these points as the \emph{closure space}. The factorization conjecture provides a complete classification of the properties of these points. The closure space is identified with the hypersurface $Z = 0$, given explicitly by
\bea
\label{pointsinfty}
X_1 X_2 \cdots X_{N} = 0.
\eea
Neglecting the $Z$ coordinate, this defines a hypersurface in $\mathbb{C}P^{N-1}$ with homogeneous coordinates $X_i$. The hypersurface consists of $N$ components, each isomorphic to $\mathbb{C}P^{N-2}$. These components intersect pairwise along $\mathbb{C}P^{N-3}$ subspaces, and the total number of pairwise intersection points (or components) is $\binom{N}{2}$. Similarly, triple intersections occur along $\mathbb{C}P^{N-4}$ subspaces, with degeneracy $\binom{N}{3}$. The general pattern of higher-order intersections follows naturally.

Therefore, the factorization conjecture implies that the points at infinity are isomorphic to $N$ copies of $\mathbb{C}P^{N-2}$. A generic point at infinity vanishes to first order and lies on one of these $\mathbb{C}P^{N-2}$ components. However, certain special points lie on the intersection of two such $\mathbb{C}P^{N-2}$ components and vanish to second order and so on. We denote by $\mathrm{ord}(x) = r$ the order of vanishing of a point $x$ lying on the hypersurface defined by~\eqref{pointsinfty}. 

The order of vanishing of a point $x$ is closely related to the notion of \emph{nilpotent Lie algebra} and it is of importance to define it explicitly. 

\begin{definition}
	A Lie algebra $\mathfrak{g}$ is called \emph{nilpotent} if its lower central series eventually terminates at the zero subalgebra, that is, if there exists $\mathcal{K}\in\mathbb{N}$ such that~\cite{serre1992lie,meurant1973introduction}
	\[
	\underbrace{[\mathfrak{g},[\mathfrak{g},\cdots,[\mathfrak{g},\mathfrak{g}]]]]]}_{\mathcal{K}~\text{times}}=\{0\}.
	\]
	Equivalently, $\mathfrak{g}$ is nilpotent if its lower central series
	\[
	\mathfrak{g}^{(0)}=\mathfrak{g}, \quad 
	\mathfrak{g}^{(i+1)}=[\mathfrak{g},\mathfrak{g}^{(i)}]
	\]
	satisfies $\mathfrak{g}^{(\mathcal{K})}=\{0\}$ for some finite $\mathcal{K}$.
\end{definition}
We call $\mathcal{K}$ the depth of the algebra. We now characterize the structure of the algebras associated to points at infinity by the following observation, which is closely tied with the factorization conjecture

\medskip
\noindent\textbf{Emergence of Nilpotent Algebras} 
Let $x$ be a point at infinity which vanishes by order $\operatorname{ord}(x)=r$. Then, the corresponding algebras associated with $x$ in the $\mathbb{Z}_N$-equivariant case is isomorphic to 
\[
\mathfrak{g} = Vir \rtimes F,
\] 
where the Virasoro subalgebra has central charge $c = 2N$, and $F$ is a nilpotent Lie algebra of depth $\mathcal{K} =N-r$.

In the following sections, we present supporting evidence for the validity of the factorization conjecture and demonstrate the emergence of nilpotent algebras for $\mathbb{Z}_2$, $\mathbb{Z}_3$, and $\mathbb{Z}_4$ within the closure spaces corresponding to the respective $\mathbb{Z}_{N}$-equivariant Virasoro algebras.

\section{$BMS_3$-like algebras corresponding to the points at infinity}
\label{closure}
As discussed in the previous sections, the moduli spaces of solutions for the $\mathbb{Z}_2$, $\mathbb{Z}_3$, and $\mathbb{Z}_4$ cases are non-compact. In~\cite{Ghazi:2025oin}, our focus was primarily on the $\mathbb{Z}_2$, $\mathbb{Z}_3$, and $\mathbb{Z}_4$ cases; however, the behavior of the solutions at infinity was not explored there. In the following subsections, we examine these examples in detail and investigate the validity of the factorization conjecture as well as its connection to the emergence of nilpotent algebras.
\subsection{The projective completion of the $\mathbb{Z}_2$}
The $\mathbb{Z}_2$ is the most simplest case. the constraints~\eqref{cons1} leads to the following identities
\bea
c^1_0=0,~~~d^1_1=0
\eea 
and $d^1_0$ and $c^1_1$ are the only nonzero coefficients in the construction of $L_n^1$. The equations~\eqref{eq1}--~\eqref{eq4} are satisfied if the following relation holds	

\bea
\label{CP11}
(d_0^1)^2+(c_1^1)^2=1,
\eea
In the absence of unitarity constraints, the parameters $d^1_0$ and $c^1_1$ can take complex values. Consequently, the space of constructions forms a cylinder, which is topologically equivalent to a sphere with two punctures.

The projective completion is then the sphere, which can be embedded into $\mathbb{C}P^2$ by defining $d^1_0 = \frac{X}{Z}$ and $c^1_1 = \frac{Y}{Z}$ as follows:
\bea
X^2 + Y^2 = Z^2.
\eea
It is then evident that the space of solutions compatible with equation~\eqref{CP11} is defined on the patch $Z \neq 0$, whereas its closure corresponds to the complementary region characterized by $Z = 0$. The above equation is completely consistent with the factorization conjecture defined by~\eqref{facconjecture} 
\bea
Z^2 = (X + iY)(X - iY).
\eea

The points at infinity are those satisfying $Z = 0$, namely
\bea
X = \pm iY.
\eea
The new constructions corresponding to~\eqref{CP11} are given by
\begin{align}
	\label{defS_n}
	L^0_n &= \frac{1}{2}\sum_q j^A_{n-q} j^A_q, \qquad n \in 2\mathbb{Z}, \\
	\label{defs_n}
	L^1_n &= \frac{d^1_0}{2}\sum_q j^A_{n-q} j^A_q + \frac{c^1_1}{2}\sum_q (-1)^q \epsilon^{AB} j^A_{n-q} j^B_q, \qquad n \in 2\mathbb{Z}+1,
\end{align}
where the standard construction is given by $d^1_0 = 1$ and $c^1_1 = 0$. However, we are interested in the algebras corresponding to the points at infinity. Since we have defined $d_0^1 = \frac{X}{Z}$ and $c^1_1 = \frac{Y}{Z}$, the limit $Z = 0$ is not well defined, and the generators $L^1_n$ become ill-defined. Nevertheless, by rescaling $F^1_n = \frac{Z}{2X} L^1_n$, everything becomes regular, and we obtain
\begin{align}
	\label{defF_n}
	L^0_n &= \frac{1}{2}\sum_q j^A_{n-q} j^A_q, \qquad n \in 2\mathbb{Z}, \\
	\label{defF}
	F^1_n &= \frac{1}{4}\sum_q j^A_{n-q} j^A_q \pm \frac{i}{4}\sum_q (-1)^q \epsilon^{AB} j^A_{n-q} j^B_q, \qquad n \in 2\mathbb{Z}+1.
\end{align}

A more intuitive way to derive the above algebras is to solve equation~\eqref{CP11} directly as follows:
\[
d^1_0 = \frac{u + u^{-1}}{2}, \qquad c_1^1 = \frac{u - u^{-1}}{2i},
\]
and then consider the limits $u \to 0$ and $u \to \infty$. Let us focus on the limit $u \to 0$. In this case, the coefficients are approximately
\[
d^1_0 \sim \frac{1}{2u}, \qquad c^1_1 \sim \frac{i}{2u},
\]
and then, by considering the rescaling $F^1_n = u L^1_n$, we recover the generators defined in~\eqref{defF}. The Virasoro algebra then transforms into the following algebras in the limit $u \to 0$:
\begin{align}
	[F^1_n, L^0_m] &= (n - m) F^1_{n+m}, \\
	[F^1_n, F^1_m] &= u^2 (n - m) L^0_{n+m} + \frac{d u^2}{12} n (n^2 - 1) \delta_{n+m,0},
\end{align}
which has a well-defined limit as $u \to 0$. Consequently, the corresponding algebra simplifies to
\begin{align}
	\label{ClosurealgebraZ2}
	[L^0_n, L^0_m] &= (n - m) L^0_{n+m} + \frac{d}{12} n (n^2 - 1) \delta_{n+m,0}, \\
	\label{Vir2'}
	[F^1_n, L^0_m] &= (n - m) F^1_{n+m}, \\
	\label{Vir3'}
	[F^1_n, F^1_m] &= 0.
\end{align}
Suppose we allow the index of generators to run over natural numbers instead of $even/odd$ separations. In that case, this is a well-known structure in mathematical physics, referred to as the Galilean Conformal Algebra (GCA)~\cite{Bagchi_2009}. And it is also the same as $BMS_3$ algebra~\cite{Barnich:2006av}.  It is worth noting that the above algebra can be generalized to include additional central extensions, where the mixed commutators in Equation~\eqref{Vir2'} can admit central terms. However, one should be careful that the above algebra has a $\mathbb{Z}_2$-equivariance structure and $n$ for $L^0_{n}$ runs over even integers and for $F^1_n$ over odd numbers. Due to this truncation, we refer to the above algebra as \emph{half Galilean Conformal Algebra} (hGCA). 

From a mathematical perspective, the algebra~\eqref{ClosurealgebraZ2}--\eqref{Vir3'} can be understood as a semidirect sum of the Virasoro algebra with an abelian algebra generated by the $F^1_n$, that is
\bea
Vir\rtimes F,
\eea
where, by defining $U_n=\frac{L^0_{2n}}{2}$\footnote{$U_0$ needs additional appropriate shifting.} and $V_n=F^1_{2n+1}$, the algebra takes the following form
\bea
\label{relabeling}
&&[\bm{U}_n, \bm{U}_m] = (n-m)\bm{U}_{n+m}+\frac{4}{12}n(n^2-1)\delta_{n+m,0},\\
&&[\bm{V}_n, \bm{U}_m] = (n+\frac{1}{2}-m)\bm{V}_{n+m},\\
&&[\bm{V}_n, \bm{V}_m] =0.
\eea
where the central charge of the theory is given by $c=4$. Since the construction of the generators in~\eqref{defF} relies explicitly on the antisymmetric tensor $\epsilon^{AB}$, one might naively expect the central charge to be $c=2$. However, the central charge of the algebras corresponds to the closure points doubles, yielding $c=4$.

From a physical perspective, it is more natural to introduce half-integer indices by setting $r=n+\tfrac{1}{2}$ and defining $G_r = V_n$. In terms of these variables, the commutator of $G_r$ with $U_m$ acquires a more transparent structure.
\bea
[U_m,G_r]=(m-r)G_{m+r},
\eea
and by comparing with the commutator of a generic primary field $\phi$ of conformal weight $h$, whose modes $\phi_n$ satisfy
\bea
[U_m,\phi_n]=\Big((h-1)m-n\Big)\phi_{n+m},
\eea
We may conclude that $G_r$ can be regarded as a primary field of conformal weight $h=2$. However, since $r \in \mathbb{Z}+\tfrac{1}{2}$, the corresponding primary field $G(z)$ necessarily belongs to the twisted sector of the theory.

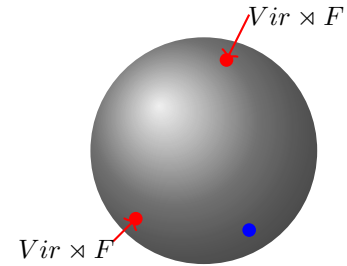
\begin{figure}[h!]
	\centering
	\begin{tikzpicture}[scale=3]
		
		\shade[ball color=white!10!gray, opacity=0.9] (0,0) circle (0.5);
		
		\fill[red] (0.1,0.4) circle (0.03);   
		\fill[red] (-0.3,-0.3) circle (0.03); 
		\fill[blue] (0.2,-0.35) circle (0.03); 
		
		\draw[->, thick, red] (0.2,0.6) -- (0.1,0.4) node[midway, above right, black] {$Vir\rtimes F$};
		\draw[->, thick, red] (-0.4,-0.4) -- (-0.3,-0.3) node[midway, below left, black] {$Vir\rtimes F$};
		
	\end{tikzpicture}
	\caption{The projective moduli space of the $\mathbb{Z}_2$ case with two points at infinity (red marked points) and one blue marked point with central charge $c=D$, which corresponds to the standard construction.}
	\label{figsphere}
\end{figure}

The general picture of the theory is now complete, which has been shown in Table~\ref{table1} and Fig~\ref{figsphere}. The projective completion of $\mathbb{Z}_2$ is divided into three distinct parts as follows.

\begin{table}[h]
	\centering
	\begin{tabular}{|c|l|}
		\hline
		\textbf{Case} & \textbf{Description} \\
		\hline
		1 & \parbox{8cm}{The Virasoro algebra corresponding to the point $c^1_1 = 0$, with central charge $c = D$ not necessarly equal to two.} \\
		\hline
		2 & \parbox{8cm}{Two closure points with algebra $Vir \rtimes F$, where $F$ is an abelian ideal and the central charge equals $c = 4$.} \\
		\hline
		3 & \parbox{8cm}{The Virasoro algebras related to the other points, with central charge $c = 2$, where $\epsilon^{ij}$ is always present.} \\
		\hline
	\end{tabular}
	\caption{the projective space of the constructions of the $Z_2$ corresponding to the algebras, diveved into three distinct part up to the equivalence relation}
	\label{table1}
\end{table}

As we saw, the Lie algebras related to the points at infinity were $g=Vir\rtimes F$, where $F$ was generated by $F^1_n$. The Lie algebra $F$ is a nilpotent Lie algebra with depth $\mathcal{K}=1$, as we have
\bea
[F^1_n,F^1_m]=0,
\eea
for all $F^1_n$. This observation is consistent with the emergence of the nilpotent algebras conjecture, since the points at infinity correspond to two isolated points, each exhibiting a first-order vanishing behavior. We examine the factorization conjecture in the next subsection.

\subsubsection{Factorization of the closure space}
The points at infinity have a simple feature. If we define $d_0^1=\tfrac{X}{Z}$ and $c_1^1=\tfrac{Y}{Z}$, then the points correspond precisely to the limit $ Z\to 0$. Geometrically, these are exactly the points that lie in the locus $Z=0$ in the projective completion with homogeneous coordinates $[X:Y:Z]$. In this regime, the defining relation~\eqref{CP11} simplifies, and in fact, factorizes as
\bea
(X+iY)(X-iY)=0,
\eea
where ingoring the $Z$ coordinate, its consist of two points $[X:\pm iX]$ embeded in $\mathbb{C}P^1$. If we take $\mathbb{C}P^0$ just as a point, the closure space is just two $\mathbb{C}P^0$. It is in agreement with the previous conjecture which stated in section~\ref{factorization}.

\subsection{The projective completion of the $\mathbb{Z}_3$}

The next step is to investigate the algebras associated with the closure points of the surfaces related to the $\mathbb{Z}_3$ case. As in the $\mathbb{Z}_2$ setting, the study of such points reveals a similarity: the closure space again exhibits a natural factorization, hinting at the existence of a universal pattern underlying the projective completions of $\mathbb{Z}_N$. This observation strongly suggests that such factorization behavior of these moduli spaces is not accidental but rather governed by algebraic principles that manifest consistently across different cyclic groups and their relation to the Virasoro algebra.

Nevertheless, there are also essential differences that distinguish the $\mathbb{Z}_3$ case from its $\mathbb{Z}_2$ counterpart. The projective completion of the $\mathbb{Z}_3$ moduli space develops higher order singularities, in contrast to the smooth completion observed for $\mathbb{Z}_2$. This subtlety plays a crucial role in shaping the associated algebraic structures. In particular, the extension of the Virasoro algebra depends sensitively the order of singularities.

For higher order singular points, the analysis shows that the Virasoro algebra extends through the addition of an abelian algebra, resembling the structure already encountered in the $\mathbb{Z}_2$ case. By contrast, for the first order singular points the situation is more intricate: the Virasoro algebra is extended by a genuinely non-abelian algebra, giving rise to a richer and less trivial structure. 

In the following subsections, we explore these aspects in detail, examining how the different singular points of the $\mathbb{Z}_3$, give rise to qualitatively different algebraic extensions, and how these fit into the broader pattern suggested by the $\mathbb{Z}_2$ case.

\subsubsection{The Moduli Space of the $\mathbb{Z}_3$-Case}
The $\mathbb{Z}_3$ case provides the next simplest example for examining the factorization conjecture and comparing the corresponding space with that of the $\mathbb{Z}_2$ case. In this instance, the structure of the solution space is richer, forming a four-dimensional surface.

The constraints~\eqref{cons1} determine the following relations between $c^k_{\omega}$ and $d^k_{\omega}$.
\bea
\label{independentcoe}
&&d^1_2=d^1_1e^{\frac{2\pi i}{3}},~~~d_2^2=d_1^2e^{\frac{4\pi i}{3}}\\
&&c_2^1=-c^1_1e^{\frac{2\pi i}{3}},~~~c^2_2=-c_1^2e^{\frac{4\pi i}{3}},~~~c^k_0=0,
\eea
Working with the independent coefficients, the consistency of Equations~\eqref{eq1}--\eqref{eq4} yields the following constraints on the coefficients.
 \begin{widetext}
	\bea
	&&d^1_0d^2_0+2(d_1^1d_1^2+c_1^1c_1^2)=1,\\
	&&d_1^1d_1^2-c_1^1c_1^2+e^{\frac{4\pi i}{3}}d^1_1d^2_0+e^{\frac{2\pi i}{3}}d^1_0d^2_1=0,\\
	&&e^{\frac{2\pi i}{3}} d^1_0c^2_1+e^{\frac{4\pi i}{3}}d^2_0c^1_1-(d_1^1c_1^2+c_1^1d_1^2)=0\\\label{Z3Eq3}
	&&(d^1_0)^2-e^{\frac{2\pi i}{3}}\Big((d^1_1)^2+(c^1_1)^2\Big)=d^2_0,\\\label{Z3Eq4}
	&&(d_1^1)^2-(c_1^1)^2-e^{\frac{2\pi i}{3}}d^1_0d^1_1=d_1^2,\\\label{Z3Eq5}
	&&-e^{\frac{2\pi i}{3}}d^1_0c^1_1-2d_1^1c_1^1=c_1^2,\\
	&&(d^2_0)^2-e^{\frac{4\pi i}{3}}\Big((d_1^2)^2+(c_1^2)^2\Big)=d^1_0,\\
	&&-e^{\frac{4\pi i}{3}}d^2_0d^2_1+(d_1^2)^2-(c_1^2)^2=d_1^1,\\
	&&-e^{\frac{4\pi i}{3}}d^2_0c^2_1-2d_1^2c_1^2=c_1^1,\\
	&&(d^1_0d^2_1+d^1_1d^2_0)(e^{\frac{4\pi i}{3}}-1)+(d^1_1d^2_1-c^1_1c^2_1)(e^{\frac{2\pi i}{3}}-e^{\frac{4\pi i}{3}})+d^1_1d^2_0(1-e^{\frac{2\pi i}{3}})=0,\\
	&&(d^1_1c^2_1+c^1_1d^2_1)(e^{\frac{2\pi i}{3}}-e^{\frac{4\pi i}{3}})-d^1_0c^2_1(e^{\frac{2\pi i}{3}}-1)-c^1_1d^2_0(1-e^{\frac{2\pi i}{3}})=0    
	\eea
\end{widetext}
The above equations seem too complicated to be solved analytically. However, we can use Equations~\eqref{Z3Eq3}--\eqref{Z3Eq5} to express $d^2_0$, $d^2_1$, and $c^2_1$ in terms of $d^1_0$, $d^1_1$, and $c^1_1$. To simplify the notation, let us define $d^1_0 \equiv x$, $d^1_1 \equiv w$, and $c^1_1 \equiv y$. Eliminating $d^2_0$, $d^2_1$, and $c^2_1$ from the equations then yields the following simplified system:
\bea
\label{smoothsolutions}
&&x^3+2w^3-3\omega x(w^2+y^2)-6wy^2=1,
\eea
Here, $\omega \equiv e^{\frac{2\pi i}{3}}$. The above equation is analogous to equation~\eqref{CP11} in the $\mathbb{Z}_2$ case. As in the case $\mathbb{Z}_2$, let us define $x=\frac{X}{Z}$, $w=\frac{W}{Z}$ and $y=\frac{Y}{Z}$, then the projective completion of the affine curve transforms into the following form
\bea Z^3 =X^3+2W^3-3\omega X(W^2+Y^2)-6WY^2,
\eea
This  is an algebraic surface embeded in $\mathbb{C}P^3$. Although there is no a priori reason to expect a factorization of the right-hand side of the above equation, it indeed occurs, and we examine this property in detail in the next subsection.

\subsubsection{Factorization of the closure space}
The space of solutions to the equation~\eqref{smoothsolutions} has closure components, which can be identified in the limit $Z \to 0$, as follows
\begin{equation}
	\label{decomposition}
	\begin{aligned}
		&F(X,W,Y,Z=0)= X^3 + 2W^3 - 3\omega X(W^2 + Y^2) - 6W Y^2 \\
		&= (\omega X + 2W)(\omega X - W + \sqrt{3}Y)(\omega X - W - \sqrt{3}Y) = 0,
	\end{aligned}
\end{equation}
which are the union of three $C\mathbb{P}^1$ in $\mathbb{C}P^2$ $[X,W,Y,Z=0]$ which is in agreement to the factorization conjecture~\ref{factorization}. The three  $C\mathbb{P}^1$ intersect pairwise; however, there is no common intersection point on the three $C\mathbb{P}^1$: 
\begin{align}
	L_1&: \ \omega X + 2W = 0, \\
	L_2&: \ \omega X - W + \sqrt{3}\, Y = 0, \\
	L_3&: \ \omega X - W - \sqrt{3}\, Y = 0,
\end{align}
And we have 
\bea
\label{singular31}
&&L_1 \cap L_2 = [2 : -\omega : -\sqrt{3}\,\omega],\\\label{singular32}
&&L_1 \cap L_3  = [2 : -\omega : \sqrt{3}\,\omega],\\\label{singular33}
&& L_2 \cap L_3  = [1 : \omega : 0],
\eea
The intersection points are the singular points of the closure space.

\subsubsection{Algebras correspond to the first order singular points}

A generic point $p$ of the closure space~\eqref{decomposition} typically lies on one of the three $\mathbb{C}P^1$ components, with a first-order vanishing, $\operatorname{ord}(p) = r = 1$. Therefore, if the conjecture of the emergence of nilpotent algebras holds, the corresponding algebras associated with these points should take the form $Vir \times F$, where $F$ is a nilpotent algebra of depth $\mathcal{K} = N - 1 = 2$. Let us now verify this property for the points defined by $X = 0$. In this limit, the curve develops three closure points, given by
\bea
W = 0,
\qquad
W = \pm \sqrt{3}Y.
\eea
The algebras associated with these closure points can be identified explicitly, and they turn out to correspond precisely to the following solutions:

\bea
&&w=0,~~~y=\frac{Y}{Z},\\
&&w=\frac{\sqrt{3}Y}{Z},~~~y=\frac{Y}{Z},\\
&&w=\frac{-\sqrt{3}Y}{Z},~~~y=\frac{Y}{Z},
\eea
These points are not belong to the intersections of three $\mathbb{C}P^1$, therefore, they are not higher order vanishing points. The corresponding algebras exhibit interesting and nontrivial structures.

\paragraph{First solution.}

For the first solution, the relevant generators take the form
\bea
&&L^1_n=\frac{Y}{2Z}\sum_q(e^{\frac{2\pi i}{3}q}-\omega e^{\frac{4\pi i}{3}q})\epsilon^{AB}j^A_{n-q}j^B_q,\\
&&L^2_n=-\frac{Y^2}{2Z^2}\sum_q(\omega+e^{\frac{2\pi i}{3}q}+\omega^2e^{\frac{4\pi i}{3}q})j^A_{n-q}j^A_q,
\eea
As expected, both operators are singular in the strict limit $Z \to 0$. To obtain a well-defined algebra, we introduce the rescaled operators, $F^1_n=\frac{Z}{Y}L^1_n$ and $F^2_n=\frac{Z^2}{Y^2}L^2_n$. In terms of these generators, together with the Virasoro operators $L^0_n$, the algebra closes consistently and takes the following form:
\bea
&&[L^0_n,L^0_m]=(n-m)L^0_{n+m}+\frac{c}{12}n(n^2-1)\delta_{n+m,0},\\
&&[L^0_n,F^k_m]=(n-m)F^k_{n+m},\\
&&[F^1_n,F^1_m]=(n-m)F^2_{n+m},\\
&&[F^2_n,F^2_m]=0,\\
&&[F^2_n,F^1_m]=0,
\eea
where $k=1,2$.

\paragraph{Second solution.}

The second solution is more complicated. This solution leads to the following generators 
\begin{align}
	L^1_n &= \frac{Y}{2Z}\sum_q \Bigl( (\sqrt{3}e^{\frac{2\pi i}{3}q} + \sqrt{3}\omega e^{\frac{4\pi i}{3}q}) j^A_{n-q} j^A_q \nonumber \\
	&\qquad\qquad + (e^{\frac{2\pi i}{3}q} - \omega e^{\frac{4\pi i}{3}q}) \epsilon^{AB} j^A_{n-q} j^B_q \Bigr), \\[1ex]
	L^2_n &= \frac{Y^2}{2Z^2}\sum_q \Bigl( (-4\omega + 2e^{\frac{2\pi i}{3}q} + 2\omega^2 e^{\frac{4\pi i}{3}q}) j^A_{n-q} j^A_q \nonumber \\
	&\qquad\qquad + 2\sqrt{3}(\omega^2 e^{\frac{4\pi i}{3}q} - e^{\frac{2\pi i}{3}q}) \epsilon^{AB} j^A_{n-q} j^B_q \Bigr),
\end{align}
As before, we define the rescaled operators, $F^1_n=\frac{Z}{Y} L^1_n$ and $F^2_n=\frac{Z^2}{Y^2} L^2_n$, in order to remove the singular behavior in the $ Z\to 0$ limit. In terms of these rescaled generators, the algebra is the same as the previous algebra:
\bea
&&[F^1_n,F^1_m]=(n-m)F^2_{n+m},\\
&&[F^1_n,F^2_m]=0,\\
&&[F^2_n,F^2_m]=0,
\eea

\paragraph{Third solution.}

The third solution has the same similarity as the second solution, with a minor difference, and leads to the following generators 
\begin{align}
	L^1_n &= \frac{Y}{2Z}\sum_q \Bigl( \bigl(-\sqrt{3}e^{\frac{2\pi i}{3}q} - \sqrt{3}\omega e^{\frac{4\pi i}{3}q}\bigr) j^A_{n-q} j^A_q \nonumber \\
	&\qquad\qquad + \bigl(e^{\frac{2\pi i}{3}q} - \omega e^{\frac{4\pi i}{3}q}\bigr) \epsilon^{AB} j^A_{n-q} j^B_q \Bigr), \\[1ex]
	L^2_n &= \frac{Y^2}{2Z^2}\sum_q \Bigl( \bigl(-4\omega + 2e^{\frac{2\pi i}{3}q} + 2\omega^2 e^{\frac{4\pi i}{3}q}\bigr) j^A_{n-q} j^A_q \nonumber \\
	&\qquad\qquad - 2\sqrt{3}\bigl(\omega^2 e^{\frac{4\pi i}{3}q} - e^{\frac{2\pi i}{3}q}\bigr) \epsilon^{AB} j^A_{n-q} j^B_q \Bigr),
\end{align}
It also leads to the same algebra as the two algebras mentioned above. 

All the algebras discussed above are isomorphic to the Lie algebra
\bea
g=Vir\rtimes F,
\eea
where the Virasoro subalgebra $Vir$ has central charge $c=6$, and $F$ is a nilpotent Lie algebra of depth $\mathcal{K}=2$, generated by the operators $F^1_n$ and $F^2_n$. These observations are all in agreement with the emergence of nilpotent algebras.

However, as we will demonstrate in the next subsection, the situation becomes more interesting for higher order singular points of the closure space, where the structure of the extension algebra differs from that of the first order singular points.

\subsubsection {Algebras Correspond to the Singular Points}

As mentioned earlier, the closure space of the $\mathbb{Z}_3$ case in~\eqref{decomposition} consists of three $\mathbb{C}P^1$ components that intersect pairwise. The corresponding intersection points are precisely given by~\eqref{singular31}--\eqref{singular33}.  Therefore, it is very interesting to explore the nature of the closure algebra related to these points for examining the emergence of nilpotent algebra and the relation between the depth and order of vanishing of the singularities. The intersection point $L_2 \cap L_3  = [1 : \omega : 0]$ leads to the following relation
\bea
(\omega X+2W)(\omega X-W)^2=0,
\eea
where $W=\omega X$ is the source of higher order singularity and is a degenerate point. The corresponding algebra is as follows
\bea
F^1_n&=&\frac{1}{2}\sum_q(1+\omega e^{\frac{2\pi i}{3}q}+\omega^2 e^{\frac{4\pi i}{3}q})j^A_{n-q}j^A_q,\\
F^2_n&=&0
\eea
which is very different from the other closure algebras, where both $F^1_n$ and $F^2_n$ contribute to the extensions of the Virasoro algebra. The corresponding algebra takes the following form
\begin{equation}
	\begin{aligned}
		[L^0_n, L^0_m] &= (n-m)L^0_{n+m} + \frac{c}{12}n(n^2-1)\delta_{n+m,0}, \\
		[L^0_n, F^1_m] &= (n-m)F^1_{n+m}, \\
		[F^1_n, F^1_m] &= 0,
	\end{aligned}
\end{equation}

Using a similar relabeling~\eqref{relabeling}, the above algebra is isomorphic to $\mathfrak{g} = \mathrm{Vir} \rtimes F$, where the Virasoro algebra has central charge $c = 6$, and $F$ is a nilpotent Lie algebra of depth $\mathcal{K} = 1$. This value coincides with $\mathcal{K} = N - r$, where $N = 3$ and $r = 2$ is the order of vanishing of the singularity. An interesting feature of the above construction is that the $F^1_n$ operators are independent of the epsilon tensor. Consequently, by allowing the index $A$ to range over $A = 1, \dots, D$, one can generalize the algebra to the $u(1)^D$ Kac--Moody algebra, rather than being restricted to $u(1)^2$, therefore, the central charge can be $c=3D$.

 The other two singular points, $[2,-\omega,\mp\sqrt{3}\omega]$, also show a similar pattern with different generators
\begin{align}
	F^1_{n\mp} &= \frac{1}{2}\sum_q \Bigl( \bigl(2 - \omega e^{\frac{2\pi i}{3}q} - \omega^2 e^{\frac{4\pi i}{3}q}\bigr) j^A_{n-q} j^A_q \nonumber \\
	&\qquad\qquad + \bigl(\mp\sqrt{3}\omega e^{\frac{2\pi i}{3}q} \pm \sqrt{3}\omega^2 e^{\frac{4\pi i}{3}q}\bigr) \epsilon^{AB} j^A_{n-q} j^B_q \Bigr), \\[1ex]
	F^2_n &= 0,
\end{align}
However, there exists a crucial difference that the central charge of the above algebras is $c=6$, as there exists the epsilon tensor in the construction.

We are now in a position to summarize the general features of the $\mathbb{Z}_3$ case:
The closure space in the $\mathbb{Z}_3$ separates into two distinct branches. For the first order vanishing points, the corresponding algebras are isomorphic to $Vir \times F$, where $F$ is a nilpotent Lie algebra of depth $\mathcal{K}=2$. The central charge in this case depends on the chosen point and takes the value $c=6$. In contrast, the higher order vanishing points exhibits a different structure. There exists precisely one singular point, namely $[1,\omega,0]$, whose associated algebra is $Vir \rtimes F$, where $F$ is a nilpotent algebra of depth $\mathcal{K}=1$ and the central charge is $c=3D$. In addition, there are two further singular points, each corresponding to the algebra $Vir \rtimes F$ with the same depth $\mathcal{K}=1$, but with central charge $c=6$.

\subsection{The projective completion of the $\mathbb{Z}_4$}

Although the analysis of the $\mathbb{Z}_2$ and $\mathbb{Z}_3$ cases provides convincing evidence for the factorization conjecture and the emergence of nilpotent algebras, examining the $\mathbb{Z}_4$ case would be instructive to further corroborate these observations. As in the $\mathbb{Z}_2$ and $\mathbb{Z}_3$ setting, the study of such points reveals a similarity: the closure space again exhibits a natural factorization. This observation strongly suggests that such factorization behavior of these moduli spaces is not accidental but rather governed by algebraic principles that manifest consistently across different cyclic groups and their relation to the Virasoro algebra.

For higher-order singular points with an order of vanishing equal to three, the analysis reveals that the Virasoro algebra extends via the addition of an abelian algebra, analogous to the structure observed in the $\mathbb{Z}_3$ case. In contrast, for lower-order singularities, the Virasoro algebra extends through a non-abelian algebra, the depth of which depends on the order of the singularity. These findings are entirely parallel to the $\mathbb{Z}_3$ case, and the purpose of these considerations is merely to provide greater confidence in the previous results.

\subsubsection{Factorization of the Closure Space}
The number of equations in the $\mathbb{Z}_4$ case is quite large and is presented in Appendix~\ref{AppA}. However, they can all be reduced to a single independent equation. The constraints~\eqref{cons1} imply that the coefficients are not independent, leading to the following relations:
\begin{equation}
	d^1_3 = i d^1_1,\quad d^1_2 = 0,\quad d^2_3 = -d^2_1,\quad d^3_3 = -i d^3_1,\quad d^3_2 = 0,
\end{equation}
\begin{equation}
	c^1_3 = -i c^1_1,\quad c^2_3 = c^2_1,\quad c^2_2 = 0,\quad c^3_3 = i c^3_1.
\end{equation}

For notational simplicity, we again set $d^1_0 \equiv x$, $d^1_1 \equiv w$, $c^1_1 \equiv y$, and introduce one additional variable $c^1_2 \equiv u$. It can then be readily verified that all the equations in~\ref{AppA} reduce to the following single equation:
\begin{align}
	\label{affine2}
	-4 w^4 &- 4 i w^2 x^2 + x^4 + 8 w^2 y^2 - 4 i x^2 y^2 - 4 y^4 \nonumber \\
	&\quad - 16 w x y u + 4 i w^2 u^2 + 2 x^2 u^2 + 4 i y^2 u^2 + u^4 = 1,
\end{align}

using the following coordinates, the factorization is clear
\begin{align}
	x_1 &= x + (1+i)w, & x_2 &= x - (1+i)w, \\
	x_3 &= (1-i)y + u, & x_4 &= (1-i)y - u
\end{align}
where the equation~\eqref{affine2} takes the compact form
\bea
\label{affine3}
(x_1+ix_3)(x_1-ix_3)(x_2+ix_4)(x_2-ix_4)=1.
\eea

The projective completion can be simply obtained by homogenizing the coordinates as $x = \frac{X}{Z}$, $w = \frac{W}{Z}$, $y = \frac{Y}{Z}$, and $u = \frac{U}{Z}$. The hyperplanes defined by
\begin{align}
	L_1&: \  X + (1+i)W+(1+i)Y+iU = 0, \\
	L_2&: \  X + (1+i)W-(1+i)Y-iU = 0, \\
	L_3&: \  X - (1+i)W+(1+i)Y-iU = 0,\\
	L_4&: \  X - (1+i)W-(1+i)Y+iU = 0.
\end{align}
Each of these hyperplanes defines a $\mathbb{C}P^2$ embedded in $\mathbb{C}P^3$ defined by the coordinates $[X:W:Y:U]$. The total number of such hyperplanes is $\binom{4}{1} = 4$, which agrees with the expectation explained in Section~\ref{Factorizationconjecture}. These $\mathbb{C}P^2$'s intersect pairwise, $L_i \cap L_j$, in $\mathbb{C}P^1$, and the total number of such pairwise intersections is $\binom{4}{2} = 6$. The triple intersections $L_i \cap L_j \cap L_k$ yield four distinct points, corresponding to $\binom{4}{3} = 4$. 

\subsection{Different singularities and the corresponding algebras }
 We provide one example for each singular point to demonstrate that the general pattern is consistent with our previous results and observations. The most singular points are the easiest to deal with. Let us choose the hyperplanes $L_1, L_2, L_3$ and consider their intersection point $L_1 \cap L_2 \cap L_3$. One can easily show that such a point has the following form:
 \begin{equation}
 	[-(1+i)W : W : W : -(1-i)W],
 \end{equation}
 with $W \neq 0$. We may safely set $W = 1$. The rescaled generator $F^1_n=ZL^1_n$ then take the following form:
\begin{align}
	F^1_n &= \frac{1}{2}\sum_q \Bigl( -(1+i) + e^{\pi i q/2} + (1-i)e^{3\pi i q/2} \Bigr) j^A_{n-q} j^A_q \nonumber \\
	&\quad + \frac{1}{2}\sum_q \Bigl( e^{\pi i q/2} - (1-i)e^{\pi i q} - i e^{3\pi i q/2} \Bigr) \epsilon^{AB} j^A_{n-q} j^B_q, \\[1ex]
	F^2_n &= F^3_n = 0,
\end{align}
where the corresponding algebra is isomorphic to $\mathrm{Vir} \rtimes F$, where the central charge of the Virasoro algebra is $c=8$ and $F$ is a the depth-one Lie algebra generated by $F^1_n$, as expected since the order of the singularity is three, giving $\mathcal{K} = 4 - 3 = 1$.

For another example, let us choose a typical point on the pairwise intersection $L_1 \cap L_2$, assuming that the point does not lie on the other two hyperplanes $L_3$ and $L_4$. A generic point of this type has the following form:
\begin{equation}
	[-(1+i)W : W : Y : -(1-i)Y], \qquad W \neq \pm Y.
\end{equation}
Let us choose $W = 1$ and $Y = 0$. The rescaled generators $F^1_n = Z L^1_n$ and $F^2_n = Z^2 L^2_n$ then take the following form:
\begin{align}
	F^1_n &= \frac{1}{2}\sum_q \Bigl( -(1+i) + e^{\pi i q/2} + (1-i)e^{3\pi i q/2} \Bigr) j^A_{n-q} j^A_q, \\
	F^2_n &= \sum_q \Bigl( i - e^{\pi i q/2} - i e^{\pi i q} + e^{3\pi i q/2} \Bigr) j^A_{n-q} j^A_q, \\
	F^3_n &= 0,
\end{align}
where, due to the presence of the epsilon tensor in the above construction, the central charge of the theory is $c = 8$ when we consider the $u(1)^2$ Kac--Moody algebra. By choosing $W = 0$ and $Y = 1$, one finds that $F^1_n$ depends entirely on the epsilon tensor, while $F^2_n$ depends only on the delta tensor. The corresponding algebra $ F$ generated by $F^1_n$ and $F^2_n$ takes the following form:
\begin{align}
	[F^1_n, F^1_m] &= (n-m) F^2_{n+m}, \\
	[F^2_n, F^2_m] &= 0, \\
	[F^2_n, F^1_m] &= 0, 
\end{align}
which is a nilpotent algebra of depth two, consistent with the order of the associated singularity. The analysis for points which are solely on a single component $L_i$ is similar. While it is straightforward (though tedious) to show that the resulting algebras are nilpotent of depth three, we will not pursue it here.

\section{Structure of New Algebras and Their Central Extensions}
\label{newalgebras}
In the previous sections, we analyzed the algebras corresponding to the points at infinity for the space of $\mathbb{Z}_N$ Virasoro constructions. As demonstrated, the general pattern of these algebras is of the form $\mathrm{Vir} \rtimes F$, where $F$ is either an abelian or a non-abelian ideal of the full algebra. In this section we ingore the standard $\mathbb{Z}_N$ gradeding and allow the index $n$ to rus over all integer numbers. As mentioned in the section~\ref{closure}, the algebras associated to the points at infinity is completely known in the literature as is just the well-known $BMS_3$ algebra
\begin{align}
	[L_n, L_m] &= (n - m) L_{n+m} + \frac{c_1}{12} n (n^2 - 1) \delta_{n+m,0}, \\
	[F_n, L_m] &= (n - m) F_{n+m}+\frac{c_2}{12} n (n^2 - 1) \delta_{n+m,0} \\
	[F_n, F_m] &= 0.
\end{align}
Due to the $\mathbb{Z}_2$ grading, $c_2$ vanishes in our case. The $BMS_3$ algebra represents the algebra of asymptotic symmetries of three-dimensional flat space. It is interesting to find the central extension of the previous algebras, which generalizes the $BMS_3$ algebra in a natural way. Consider the set of generators $F^i_n$ with $i = 0, \dots, N-1$. The central extension of the previous algebras then takes the following form:
\begin{equation}
	[F^i_n, F^j_m] = (n-m)\alpha^{ij} F^{i+j}_{n+m} + \frac{A^{ij}}{12} n(n^2-1) \delta_{n+m,0},
\end{equation}
where $A^{ij}$ is a symmetric matrix and $\alpha^{ij}$ is defined by
\begin{equation}
	\alpha^{ij} = 
	\begin{cases}
		1 & \text{if } i + j < N, \\
		0 & \text{if } i + j \geq N.
	\end{cases}
\end{equation}
In the above construction, we identify $L_n \equiv F^0_n$. The commutation relation is consistent with that of primary fields of conformal weight $h = 2$, however the novel part of our work is that the above algebras can be constructed through the $u(1)^2$ Kac-Moody algebra. Aside from the $\mathbb{Z}_2$ case, the geometric interpretation of the above algebras remains unclear (to the best of the authors' knowledge). It is an interesting possibility if such an algebras arises from the asymptotic symmetries of three dimensional space-times or similar constructions. 

\section{Conclusion}
The $\mathbb{Z}_N$-equivariant Virasoro algebra was introduced in~\cite{Ghazi:2025oin}. It was shown that, for a fixed Kac–Moody algebra associated with the group $U(1)^2$, there exist infinitely many distinct realizations of the Virasoro generators $L_n$ in terms of current modes $j^A_n$ beyond the standard Sugawara construction. Specifically, the Virasoro generators $L_n$ are decomposed into a set ${L_n^k}$, where in $L_n^k$ the mode index $n$ takes values in $n \in N\mathbb{Z} + k$. In general, the space of such constructions is isomorphic to the set of points of an $N$-dimensional algebraic variety embedded in $\mathbb{C}^N$. These algebraic varieties are, in general, non-compact. In this paper, we investigate the structure of the algebras associated with the points at infinity of these varieties. By adjoining all such points, one naturally obtains a projective variety. It is conjectured that this projective variety exhibits a form of factorization; more precisely, it can be embedded into $\mathbb{C}P^N$ in the form $Z^N=X_1X_2...X_N$ for a suitable choice of coordinates. We refer to this embedding as the factorization conjecture, and we verify its validity for $\mathbb{Z}_2$, $\mathbb{Z}_3$, and $\mathbb{Z}_4$, providing strong evidence that it holds for general $\mathbb{Z}_N$. We further show that the validity of the factorization conjecture completely determines the structure of the points at infinity (the closure space). Specifically, it consists of $N$ copies of $\mathbb{C}P^{N-2}$ that intersect pairwise along $\mathbb{C}P^{N-3}$, with a total of $\binom{N}{2}$ such components. These copies also exhibit triple intersections along $\mathbb{C}P^{N-4}$, with a total of $\binom{N}{3}$ such components, and the pattern of higher-order intersections follows naturally from this structure.

We show that the algebras corresponding to the points at infinity can generally be expressed in the form $g = Vir \rtimes F$, where $F$ is a nilpotent algebra whose depth is determined by the order of the singularity of the corresponding point. We explicitly verify this structure for $\mathbb{Z}_2$, $\mathbb{Z}_3$, and $\mathbb{Z}_4$. Moreover, we observe a clear pattern and formulate a mild conjecture: the depth $\mathcal{K}$ of the nilpotent algebra $F$ associated with a point $u$ at infinity is related to the order of its singularity, $ord(u) = r$, via $\mathcal{K}=N-r$. We establish that the algebras of the form $g = Vir \rtimes F$ are closely related to extensions of the $BMS_3$ or conformal Galilean algebra. In particular, the $\mathbb{Z}_2$ case corresponds to a truncation of the $BMS_3$ or conformal Galilean algebra, where $F$ forms an abelian ideal of the full algebra. For higher values of $N$, the structure becomes richer, incorporating extensions by non-abelian ideals. At the end, by relaxing the $\mathbb{Z}_N$-equivariance condition, we examined these algebras from a purely algebraic viewpoint and explored their possible central extensions. To the best of the authors’ knowledge, such algebras have not been previously discussed through the current algebras in the literature and represent new structures of potential interest in the broader study of infinite-dimensional symmetries.

\appendix
\section{The $\mathbb{Z}_4$-equivariant Virasoro equations}
\label{AppA}
In this appendix, we present the equations~\eqref{eq1}--\eqref{eq4} explicitly, as they were manipulated using \texttt{Mathematica}. The equations have also been simplified by incorporating the constraint~\eqref{cons1}. 

\begin{widetext}

	\begin{equation}
		\begin{aligned}
			&(c_2^1)^2 + (d_0^1)^2 = d_0^2 \\
			&(-1 - i) c_1^1 c_2^1 + (1 - i) d_0^1 d_1^1 = d_1^2 \\
			&(1 - i) c_1^1 d_0^1 - (1 + i) c_2^1 d_1^1 = c_1^2 \\
			&2i \left((c_1^1)^2 - (d_1^1)^2\right)= d_2^2 \\
			&(1 + i) c_1^1 c_2^1 - (1 - i) d_0^1 d_1^1 = -d_1^2 \\
			&(1 - i) c_1^1 d_0^1 - (1 + i) c_2^1 d_1^1 = c_1^2 \\
			&d_0^1 d_0^2 - \left(\frac{1}{2} - \frac{i}{2}\right) \left(-c_1^1 c_1^2 - d_1^1 d_1^2\right) - \left(\frac{1}{2} + \frac{i}{2}\right) \left(i c_1^1 c_1^2 + i d_1^1 d_1^2\right) = d_0^3 \\
			&\left(-\frac{1}{2} + \frac{i}{2}\right) \left(c_1^2 d_1^1 - c_1^1 d_1^2\right) - \left(\frac{1}{2} + \frac{i}{2}\right) \left(i c_1^2 d_1^1 - i c_1^1 d_1^2\right) = 0 \\
			&\left(-\frac{1}{2} - \frac{i}{2}\right) c_2^1 c_1^2 + \left(\frac{1}{2} - \frac{i}{2}\right) d_0^1 d_1^2 - i d_1^1 d_2^2 = d_1^3 \\
			&\left(\frac{1}{2} - \frac{i}{2}\right) c_1^2 d_0^1 - \left(\frac{1}{2} + \frac{i}{2}\right) c_2^1 d_1^2 + i c_1^1 d_2^2 = c_1^3 \\
			&\left(-\frac{1}{2} + \frac{i}{2}\right) \left(i c_1^1 c_1^2 - i d_1^1 d_1^2\right) - \left(\frac{1}{2} + \frac{i}{2}\right) \left(-c_1^1 c_1^2 + d_1^1 d_1^2\right) = 0 \\
			&c_2^1 d_0^2 - \left(\frac{1}{2} - \frac{i}{2}\right) \left(i c_1^2 d_1^1 + i c_1^1 d_1^2\right) - \left(\frac{1}{2} + \frac{i}{2}\right) \left(c_1^2 d_1^1 + c_1^1 d_1^2\right) = c_2^3 \\
			&\left(-\frac{1}{2} + \frac{i}{2}\right) c_2^1 c_1^2 - \left(\frac{1}{2} + \frac{i}{2}\right) d_0^1 d_1^2 - d_1^1 d_2^2 = -i d_1^3 \\
			&\left(\frac{1}{2} + \frac{i}{2}\right) c_1^2 d_0^1 + \left(\frac{1}{2} - \frac{i}{2}\right) c_2^1 d_1^2 - c_1^1 d_2^2 = i c_1^3 \\
			&c_2^1 c_2^3 + d_0^1 d_0^3 + i \left(-i c_1^1 c_1^3 - i d_1^1 d_1^3\right) - i \left(i c_1^1 c_1^3 + i d_1^1 d_1^3\right) = 1 \\
			&\left(\frac{1}{2} + \frac{i}{2}\right) c_2^1 c_1^3 + \left(\frac{1}{2} - \frac{i}{2}\right) c_1^1 c_2^3 + \left(\frac{1}{2} + \frac{i}{2}\right) d_1^1 d_0^3 + \left(\frac{1}{2} - \frac{i}{2}\right) d_0^1 d_1^3 = 0 \\
			&\left(\frac{1}{2} - \frac{i}{2}\right) c_1^3 d_0^1 + \left(\frac{1}{2} - \frac{i}{2}\right) c_2^3 d_1^1 + \left(\frac{1}{2} + \frac{i}{2}\right) c_1^1 d_0^3 + \left(\frac{1}{2} + \frac{i}{2}\right) c_2^1 d_1^3 = 0 \\
			&\left(\frac{1}{2} + \frac{i}{2}\right) c_2^1 c_1^3 + \left(\frac{1}{2} - \frac{i}{2}\right) c_1^1 c_2^3 + \left(\frac{1}{2} + \frac{i}{2}\right) d_1^1 d_0^3 + \left(\frac{1}{2} - \frac{i}{2}\right) d_0^1 d_1^3 = 0 \\
			&\left(-\frac{1}{2} + \frac{i}{2}\right) c_1^3 d_0^1 - \left(\frac{1}{2} - \frac{i}{2}\right) c_2^3 d_1^1 - \left(\frac{1}{2} + \frac{i}{2}\right) c_1^1 d_0^3 - \left(\frac{1}{2} + \frac{i}{2}\right) c_2^1 d_1^3 = 0 \\
			&2 (c_1^2)^2 + (d_0^2)^2 + 2 (d_1^2)^2 + (d_2^2)^2 = 1 \\
			&2 (c_1^2)^2 - 2 (d_1^2)^2 + 2 d_0^2 d_2^2 = 0 \\
			&d_0^2 d_0^3 - \left(\frac{1}{2} + \frac{i}{2}\right) \left(-c_1^2 c_1^3 - d_1^2 d_1^3\right) - \left(\frac{1}{2} - \frac{i}{2}\right) \left(-i c_1^2 c_1^3 - i d_1^2 d_1^3\right) = d_0^1 \\
		\end{aligned}
	\end{equation}
\end{widetext}
\begin{widetext}

	\begin{equation}
		\begin{aligned}
			&\left(-\frac{1}{2} + \frac{i}{2}\right) \left(i c_1^3 d_1^2 - i c_1^2 d_1^3\right) - \left(\frac{1}{2} + \frac{i}{2}\right) \left(-c_1^3 d_1^2 + c_1^2 d_1^3\right) = 0 \\
			&\left(-\frac{1}{2} + \frac{i}{2}\right) c_1^2 c_2^3 + \left(\frac{1}{2} + \frac{i}{2}\right) d_1^2 d_0^3 + i d_2^2 d_1^3 = d_1^1 \\
			&\left(-\frac{1}{2} + \frac{i}{2}\right) c_2^3 d_1^2 - i c_1^3 d_2^2 + \left(\frac{1}{2} + \frac{i}{2}\right) c_1^2 d_0^3 = c_1^1 \\
			&\left(-\frac{1}{2} - \frac{i}{2}\right) \left(-i c_1^2 c_1^3 + i d_1^2 d_1^3\right) - \left(\frac{1}{2} - \frac{i}{2}\right) \left(-c_1^2 c_1^3 + d_1^2 d_1^3\right) = 0 \\
			&c_2^3 d_0^2 - \left(\frac{1}{2} + \frac{i}{2}\right) \left(-i c_1^3 d_1^2 - i c_1^2 d_1^3\right) - \left(\frac{1}{2} - \frac{i}{2}\right) \left(c_1^3 d_1^2 + c_1^2 d_1^3\right) = c_2^1 \\
			&\left(-\frac{1}{2} - \frac{i}{2}\right) c_1^2 c_2^3 - \left(\frac{1}{2} - \frac{i}{2}\right) d_1^2 d_0^3 - d_2^2 d_1^3 = i d_1^1 \\
			&\left(\frac{1}{2} + \frac{i}{2}\right) c_2^3 d_1^2 - c_1^3 d_2^2 + \left(\frac{1}{2} - \frac{i}{2}\right) c_1^2 d_0^3 = -i c_1^1 \\
			&(c_2^3)^2 + (d_0^3)^2 = d_0^2 \\
			&(-1 + i) c_1^3 c_2^3 + (1 + i) d_0^3 d_1^3 = d_1^2 \\
			&(1 + i) c_1^3 d_0^3 - (1 - i) c_2^3 d_1^3 = c_1^2 \\
			&-i \left((c_1^3)^2 - (d_1^3)^2\right) + i \left(-(c_1^3)^2 + (d_1^3)^2\right) = d_2^2 \\
			&(1 - i) c_1^3 c_2^3 - (1 + i) d_0^3 d_1^3 = -d_1^2 \\
			&(1 + i) c_1^3 d_0^3 - (1 - i) c_2^3 d_1^3 = c_1^2\\
			&\left(\frac{1}{2}
			- \frac{i}{2}\right) c_2^1 c_1^2 + d_1^1 d_0^2 - \left(\frac{1}{2} + \frac{i}{2}\right) d_0^1 d_1^2 = 0 \\
			&\left(-\frac{1}{2} - \frac{i}{2}\right) c_1^2 d_0^1 + c_1^1 d_0^2 + \left(\frac{1}{2} - \frac{i}{2}\right) c_2^1 d_1^2 = 0 \\
			&\left(\frac{1}{2} + \frac{i}{2}\right) \left(i c_1^1 c_1^2 - i d_1^1 d_1^2\right) + \left(\frac{1}{2} - \frac{i}{2}\right) \left(-c_1^1 c_1^2 + d_1^1 d_1^2\right) - d_0^1 d_2^2 = 0 \\
			&\left(\frac{1}{2} + \frac{i}{2}\right) \left(i c_1^2 d_1^1 + i c_1^1 d_1^2\right) + \left(\frac{1}{2} - \frac{i}{2}\right) \left(c_1^2 d_1^1 + c_1^1 d_1^2\right) = 0 \\
			&\left(\frac{1}{2} + \frac{i}{2}\right) c_2^1 c_1^2 + i d_1^1 d_0^2 + \left(\frac{1}{2} - \frac{i}{2}\right) d_0^1 d_1^2 = 0 \\
			&\left(-\frac{1}{2} + \frac{i}{2}\right) c_1^2 d_0^1 - i c_1^1 d_0^2 - \left(\frac{1}{2} + \frac{i}{2}\right) c_2^1 d_1^2 = 0 \\
			&\left(\frac{1}{2} - \frac{i}{2}\right) c_2^1 c_1^3 - \left(\frac{1}{2} + \frac{i}{2}\right) c_1^1 c_2^3 + \left(\frac{1}{2} - \frac{i}{2}\right) d_1^1 d_0^3 - \left(\frac{1}{2} + \frac{i}{2}\right) d_0^1 d_1^3 = 0 \\
			&\left(-\frac{1}{2} - \frac{i}{2}\right) c_1^3 d_0^1 - \left(\frac{1}{2} + \frac{i}{2}\right) c_2^3 d_1^1 + \left(\frac{1}{2} - \frac{i}{2}\right) c_1^1 d_0^3 + \left(\frac{1}{2} - \frac{i}{2}\right) c_2^1 d_1^3 = 0 \\
			&-c_2^3 d_0^1 + c_2^1 d_0^3 + i \left(-c_1^3 d_1^1 - c_1^1 d_1^3\right) - i \left(c_1^3 d_1^1 + c_1^1 d_1^3\right) = 0 \\
			&\left(\frac{1}{2} + \frac{i}{2}\right) \left(c_1^2 d_1^1 - c_1^1 d_1^2\right) + \left(\frac{1}{2} - \frac{i}{2}\right) \left(i c_1^2 d_1^1 - i c_1^1 d_1^2\right) - c_2^1 d_2^2 = 0 \\
			&\left(\frac{1}{2} + \frac{i}{2}\right) \left(-c_1^1 c_1^2 - d_1^1 d_1^2\right) + \left(\frac{1}{2} - \frac{i}{2}\right) \left(i c_1^1 c_1^2 + i d_1^1 d_1^2\right) = 0 \\
		\end{aligned}
	\end{equation}
\end{widetext}	
\begin{widetext}

	\begin{equation}
		\begin{aligned}
			&\left(-\frac{1}{2} + \frac{i}{2}\right) c_2^1 c_1^3 + \left(\frac{1}{2} + \frac{i}{2}\right) c_1^1 c_2^3 - \left(\frac{1}{2} - \frac{i}{2}\right) d_1^1 d_0^3 + \left(\frac{1}{2} + \frac{i}{2}\right) d_0^1 d_1^3 = 0 \\
			&\left(-\frac{1}{2} - \frac{i}{2}\right) c_1^3 d_0^1 - \left(\frac{1}{2} + \frac{i}{2}\right) c_2^3 d_1^1 + \left(\frac{1}{2} - \frac{i}{2}\right) c_1^1 d_0^3 + \left(\frac{1}{2} - \frac{i}{2}\right) c_2^1 d_1^3 = 0 \\
			&\left(-\frac{1}{2} - \frac{i}{2}\right) \left(-c_1^1 c_1^2 - d_1^1 d_1^2\right) - \left(\frac{1}{2} - \frac{i}{2}\right) \left(i c_1^1 c_1^2 + i d_1^1 d_1^2\right) = 0 \\
			&\left(-\frac{1}{2} - \frac{i}{2}\right) \left(c_1^2 d_1^1 - c_1^1 d_1^2\right) - \left(\frac{1}{2} - \frac{i}{2}\right) \left(i c_1^2 d_1^1 - i c_1^1 d_1^2\right) + c_2^1 d_2^2 = 0 \\
			&\left(-\frac{1}{2} + \frac{i}{2}\right) c_2^1 c_1^2 - d_1^1 d_0^2 + \left(\frac{1}{2} + \frac{i}{2}\right) d_0^1 d_1^2 = 0 \\
			&\left(\frac{1}{2} + \frac{i}{2}\right) c_1^2 d_0^1 - c_1^1 d_0^2 - \left(\frac{1}{2} - \frac{i}{2}\right) c_2^1 d_1^2 = 0 \\
			&\left(-\frac{1}{2} - \frac{i}{2}\right) \left(i c_1^1 c_1^2 - i d_1^1 d_1^2\right) - \left(\frac{1}{2} - \frac{i}{2}\right) \left(-c_1^1 c_1^2 + d_1^1 d_1^2\right) + d_0^1 d_2^2 = 0 \\
			&\left(-\frac{1}{2} - \frac{i}{2}\right) \left(i c_1^2 d_1^1 + i c_1^1 d_1^2\right) - \left(\frac{1}{2} - \frac{i}{2}\right) \left(c_1^2 d_1^1 + c_1^1 d_1^2\right) = 0 \\
			&\left(-\frac{1}{2} - \frac{i}{2}\right) c_2^1 c_1^2 - i d_1^1 d_0^2 - \left(\frac{1}{2} - \frac{i}{2}\right) d_0^1 d_1^2 = 0 \\
			&\left(\frac{1}{2} - \frac{i}{2}\right) c_1^2 d_0^1 + i c_1^1 d_0^2 + \left(\frac{1}{2} + \frac{i}{2}\right) c_2^1 d_1^2 = 0 \\
		\end{aligned}
	\end{equation}
\end{widetext}

\begin{widetext}

	\begin{equation}
		\begin{aligned}	
			&\left(-\frac{1}{2} + \frac{i}{2}\right) \left(-c_1^2 c_1^3 - d_1^2 d_1^3\right) - \left(\frac{1}{2} + \frac{i}{2}\right) \left(-i c_1^2 c_1^3 - i d_1^2 d_1^3\right) = 0 \\
			&c_2^3 d_2^2 - \left(\frac{1}{2} + \frac{i}{2}\right) \left(i c_1^3 d_1^2 - i c_1^2 d_1^3\right) - \left(\frac{1}{2} - \frac{i}{2}\right) \left(-c_1^3 d_1^2 + c_1^2 d_1^3\right) = 0 \\
			&\left(-\frac{1}{2} - \frac{i}{2}\right) c_1^2 c_2^3 + \left(\frac{1}{2} - \frac{i}{2}\right) d_1^2 d_0^3 - d_0^2 d_1^3 = 0 \\
			&-c_1^3 d_0^2 - \left(\frac{1}{2} + \frac{i}{2}\right) c_2^3 d_1^2 + \left(\frac{1}{2} - \frac{i}{2}\right) c_1^2 d_0^3 = 0 \\
			&d_2^2 d_0^3 - \left(\frac{1}{2} - \frac{i}{2}\right) \left(-i c_1^2 c_1^3 + i d_1^2 d_1^3\right) - \left(\frac{1}{2} + \frac{i}{2}\right) \left(-c_1^2 c_1^3 + d_1^2 d_1^3\right) = 0 \\
			&\left(-\frac{1}{2} + \frac{i}{2}\right) \left(-i c_1^3 d_1^2 - i c_1^2 d_1^3\right) - \left(\frac{1}{2} + \frac{i}{2}\right) \left(c_1^3 d_1^2 + c_1^2 d_1^3\right) = 0 \\
			&\left(-\frac{1}{2} + \frac{i}{2}\right) c_1^2 c_2^3 - \left(\frac{1}{2} + \frac{i}{2}\right) d_1^2 d_0^3 + i d_0^2 d_1^3 = 0 \\
			&-i c_1^3 d_0^2 + \left(\frac{1}{2} - \frac{i}{2}\right) c_2^3 d_1^2 + \left(\frac{1}{2} + \frac{i}{2}\right) c_1^2 d_0^3 = 0 \\
			&\left(-\frac{1}{2} + \frac{i}{2}\right) c_2^1 c_1^3 + \left(\frac{1}{2} + \frac{i}{2}\right) c_1^1 c_2^3 - \left(\frac{1}{2} - \frac{i}{2}\right) d_1^1 d_0^3 + \left(\frac{1}{2} + \frac{i}{2}\right) d_0^1 d_1^3 = 0 \\
			&\left(\frac{1}{2} + \frac{i}{2}\right) c_1^3 d_0^1 + \left(\frac{1}{2} + \frac{i}{2}\right) c_2^3 d_1^1 - \left(\frac{1}{2} - \frac{i}{2}\right) c_1^1 d_0^3 - \left(\frac{1}{2} - \frac{i}{2}\right) c_2^1 d_1^3 = 0 \\
			&c_2^3 d_0^1 - c_2^1 d_0^3 - i \left(-c_1^3 d_1^1 - c_1^1 d_1^3\right) + i \left(c_1^3 d_1^1 + c_1^1 d_1^3\right) = 0 \\
			&\left(\frac{1}{2} - \frac{i}{2}\right) c_2^1 c_1^3 - \left(\frac{1}{2} + \frac{i}{2}\right) c_1^1 c_2^3 + \left(\frac{1}{2} - \frac{i}{2}\right) d_1^1 d_0^3 - \left(\frac{1}{2} + \frac{i}{2}\right) d_0^1 d_1^3 = 0 \\
			&\left(\frac{1}{2} + \frac{i}{2}\right) c_1^3 d_0^1 + \left(\frac{1}{2} + \frac{i}{2}\right) c_2^3 d_1^1 - \left(\frac{1}{2} - \frac{i}{2}\right) c_1^1 d_0^3 - \left(\frac{1}{2} - \frac{i}{2}\right) c_2^1 d_1^3 = 0 \\
			&\left(\frac{1}{2} - \frac{i}{2}\right) \left(-c_1^2 c_1^3 - d_1^2 d_1^3\right) + \left(\frac{1}{2} + \frac{i}{2}\right) \left(-i c_1^2 c_1^3 - i d_1^2 d_1^3\right) = 0 \\
			&-c_2^3 d_2^2 + \left(\frac{1}{2} + \frac{i}{2}\right) \left(i c_1^3 d_1^2 - i c_1^2 d_1^3\right) + \left(\frac{1}{2} - \frac{i}{2}\right) \left(-c_1^3 d_1^2 + c_1^2 d_1^3\right) = 0 \\
			&\left(\frac{1}{2} + \frac{i}{2}\right) c_1^2 c_2^3 - \left(\frac{1}{2} - \frac{i}{2}\right) d_1^2 d_0^3 + d_0^2 d_1^3 = 0 \\
			&c_1^3 d_0^2 + \left(\frac{1}{2} + \frac{i}{2}\right) c_2^3 d_1^2 - \left(\frac{1}{2} - \frac{i}{2}\right) c_1^2 d_0^3 = 0 \\
			&-d_2^2 d_0^3 + \left(\frac{1}{2} - \frac{i}{2}\right) \left(-i c_1^2 c_1^3 + i d_1^2 d_1^3\right) + \left(\frac{1}{2} + \frac{i}{2}\right) \left(-c_1^2 c_1^3 + d_1^2 d_1^3\right) = 0 \\
			&\left(\frac{1}{2} - \frac{i}{2}\right) \left(-i c_1^3 d_1^2 - i c_1^2 d_1^3\right) + \left(\frac{1}{2} + \frac{i}{2}\right) \left(c_1^3 d_1^2 + c_1^2 d_1^3\right) = 0 \\
			&\left(\frac{1}{2} - \frac{i}{2}\right) c_1^2 c_2^3 + \left(\frac{1}{2} + \frac{i}{2}\right) d_1^2 d_0^3 - i d_0^2 d_1^3 = 0 \\
			&i c_1^3 d_0^2 - \left(\frac{1}{2} - \frac{i}{2}\right) c_2^3 d_1^2 - \left(\frac{1}{2} + \frac{i}{2}\right) c_1^2 d_0^3 = 0
		\end{aligned}
	\end{equation}
\end{widetext}

	\bibliographystyle{apsrev4-1}
	\bibliography{mybib}  
\end{document}